# UV and X-ray Monitoring of AG Draconis During the 1994/1995 Outbursts[*]

J. Greiner[1], K. Bickert[1], R. Luthardt[2], R. Viotti[3], A. Altamore[4],
R. González-Riestra[5][**]

[1] Max-Planck-Institut für Extraterrestrische Physik, 85740 Garching, Germany
[2] Sternwarte Sonneberg, 96515 Sonneberg, Germany
[3] Istituto di Astrofisica Spaziale, CNR, Via Enrico Fermi 21, 00044 Frascati, Italy
[4] Dipartimento di Fisica E. Amaldi, Università Roma III, 00146 Roma, Italy
[5] IUE Observatory, ESA, Villafranca del Castillo, 28080 Madrid, Spain

**Abstract.** The recent 1994-1995 active phase of AG Draconis has given us for the first time the opportunity to follow the full X-ray behaviour of a symbiotic star during two successive outbursts and to compare with its quiescence X-ray emission. With *ROSAT* observations we have discovered a remarkable decrease of the X-ray flux during both optical maxima, followed by a gradual recovering to the pre-outburst flux. In the UV the events were characterized by a large increase of the emission line and continuum fluxes, comparable to the behaviour of AG Dra during the 1980-81 active phase. The anticorrelation of X-ray/UV flux and optical brightness evolution is shown to very likely be due to a temperature decrease of the hot component. Such a temperature decrease could be produced by an increased mass transfer to the burning compact object, causing it to slowly expand to about twice its original size.

## 1 Introduction

The symbiotic star AG Draconis (BD +67°922) plays an outstanding role inside the group of symbiotic stars (binary systems consisting of a cool, luminous visual primary and a hot compact object (white dwarf, subdwarf) as secondary component) because of its high galactic latitude ($b^{II}$=+41°), its large radial velocity of $v_r$=−148 km/s and its relatively early spectral type (K). AG Dra is probably a metal poor symbiotic binary in the galactic halo.

The historical light curve of AG Dra is characterized by a sequence of active and quiescent phases (e.g. Robinson 1969) The activity is represented by 1–2 mag light maxima (currently called *outbursts* or *eruptions*) frequently followed by one or more secondary maxima. Between the active phases AG Dra is spending long periods (few years to decades) at minimum light with small (0.1 mag) semiregular photometric variations in B and V with pseudo-periods of 300-400 days (Luthardt 1990). However, in the U band regular variations with amplitudes of 1 mag and a period of 554 days have been discovered by Meinunger (1979).

This periodicity is associated with the orbital motion of the system, as confirmed by the radial velocity observations of Kenyon and Garcia (1986).

The optical spectrum of AG Dra is typical of a symbiotic star, with a probably stable cool component which dominates the yellow-red region, and a largely variable "nebular" component with a strong blue-ultraviolet continuum and a rich emission line spectrum (e.g. Boyarchuk 1966). According to most authors the cool component is a K3 giant, which together with its large radial velocity and high galactic latitude, would place AG Dra in the halo population at a distance of about 1.2 kpc, 0.8 kpc above the galactic plane. More recently, Huang et al. (1994) suggested that the cool component may be of spectral type K0Ib, which would place it at a distance of about 10 kpc, at z = 6.6 kpc.

The UV continuum and line flux is largely variable with the star's activity. Viotti et al. (1984) studied the IUE spectra of AG Dra during the major 1980–1983 active phase, and found that the outburst was most energetic in the ultraviolet with an overall rise of about a factor 10 in the continuum, much larger than in the visual, and of a factor 2–5 in the emission line flux.

First X-ray observations of AG Dra during the quiescent phase with *Einstein* before the 1981-1985 series of eruptions revealed a soft spectrum (Anderson et al. 1981). The data are consistent with a blackbody source of kT=0.016 keV (Kenyon 1988) in addition to the bremsstrahlung source (kT=0.1 keV) suggested by Anderson et al. (1981). *EXOSAT* was pointed on AG Dra four times during the 1985–86 minor active phase, which was characterized by two light maxima in February 1985 and January 1986. These observations revealed a large X-ray fading with respect to quiescence (Piro 1986), the source being at least 5–6 times weaker in the EXOSAT thin Lexan filter in March 1985, and not detected in February 1986 (Viotti et al. 1995).

In June/July 1994 AG Dra went into a major outburst (Graslo et al. 1994), after which it gradually declined to the quiescent level in November 1994. Like the 1981–82 and 1985–86 episodes, AG Dra underwent a secondary outburst in July 1995. Here, we use all available *ROSAT* data to document the X-ray light curve of AG Dra over the past 5 years and report on the results of the coordinated *ROSAT/IUE* campaign during the 1994/1995 outbursts.

## 2  IUE Observations

AG Dra was observed by IUE as a Target of Opportunity starting on June 29, 1994. Observations have continued until February 1996, covering therefore the 1994 and 1995 outbursts as well as the period between them and the return to quiescence. In general, priority was given to low resolution spectra, but high resolution data were also obtained on most of the dates. For most of the low resolution images, both IUE apertures were used in order to have in the Large Aperture the continuum and the emission lines well exposed and in the Small Aperture (not photometric) the strongest emission lines (essentially HeII 1640 Å) not saturated. The fluxes from the Small Aperture spectra were corrected for the smaller aperture transmission by comparison with the non-saturated parts of the Large Aperture data. Some representative IUE spectra are shown in Fig. 1.

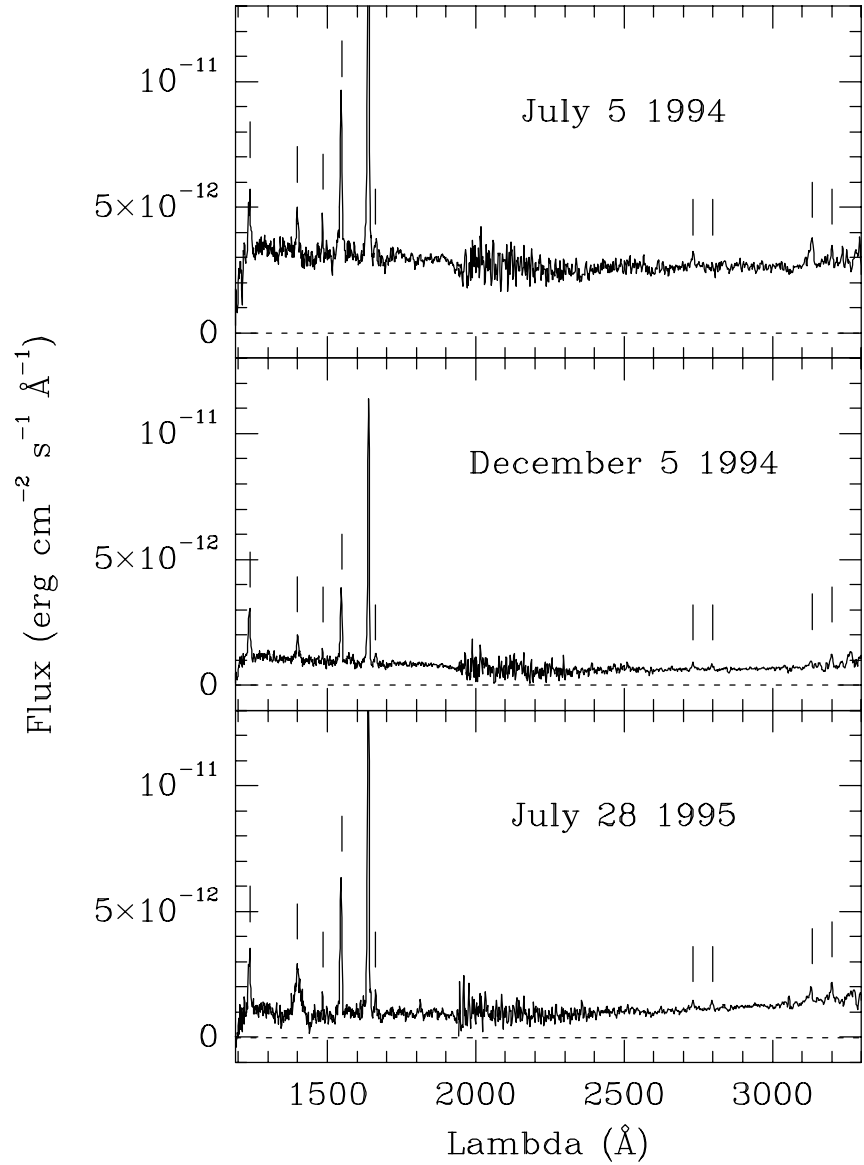

**Fig. 1.** IUE spectra of AG Dra at three different dates during the 1994-1995 activity phase: during the 1994 outburst (top), during the "quiescent" state between the outbursts (middle) and during the second outburst (bottom). The strongest line in the three spectra is HeII 1640 Å. The lines marked in the spectra are: NV 1240 Å, the blend SiIV + OIV] 1400 Å, NIV] 1486 Å, CIV 1550 Å, OIII] 1663 Å, HeII 2733 Å, MgII 2800 Å, OIII 3133 Å and HeII 3202 Å.

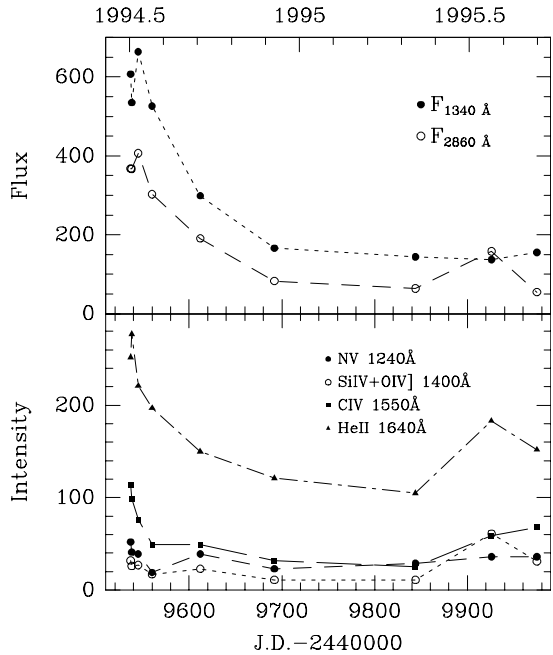

**Fig. 2.** Evolution of the UV continuum and emission lines of AG Dra during the 1994-1995 outburst. Units of continuum fluxes are $10^{-14}$ erg cm$^{-2}$ s$^{-1}$ Å$^{-1}$, and of line intensities $10^{-12}$ erg cm$^{-2}$ s$^{-1}$. All the measurements are corrected for an interstellar reddening of E(B-V)=0.06.

The behaviour of the continuum and line intensities is shown in Fig. 2. All the fluxes are corrected for an interstellar absorption of E(B-V)=0.06 (Viotti et al. 1983). The feature at 1400 Å is a blend, unresolved at low resolution of the Si IV doublet at 1393.73 and 1402.73 Å, the OIV] multiplet (1399.77, 1401.16, 1404.81 and 1407.39 Å), and S IV] 1402.73 Å. O IV]1401.16 is the dominant contributor to the blend. Because of the variable intensity of the lines, at low resolution the 1400 Å feature shows extended wings with variable extension and strength (Fig. 1). The two continuum regions were chosen, as in previous works, for being less affected by the emission lines, and for being the 1340 Å respresentative of the continuum of the hot source (nearly the Rayleigh-Jeans tail of a $10^5$ K blackbody), and the 2860 Å region representative of the HII Balmer continuum emission (see for instance Fernández-Castro et al. 1995).

## 3  ROSAT Observations

### 3.1  Observational Details

**All-Sky-Survey** AG Dra was scanned during the All-Sky-Survey over a time span of 10 days. The total observation time resulting from 95 individual scans adds up to 2.0 ksec.

**Pointed Observations** Several dedicated pointings on AG Dra have been performed in 1992 and 1993 with the ROSAT PSPC. All pointings were performed with the target on-axis. During the last *ROSAT* observation of AG Dra with the PSPC in the focal plane (already as a TOO) the Boron filter was erroneously left in front of the PSPC after a scheduled calibration observation.

When AG Dra was reported to go into outburst (Granslo et al. 1994) we immediately proposed for a target of opportunity observation (TOO) with ROSAT. AG Dra was scheduled to be observed during the last week of regular PSPC observations on July 7, 1994, but due to star tracker problems no photons were collected. For all the later *ROSAT* observations only the HRI could be used after the PSPC gas has been almost completely exhausted. Consequently, no spectral information is available for these observations. The first HRI observation took place on August 28, 1994, about 4 weeks after the optical maximum.

### 3.2 The X-ray Lightcurve of AG Dra

The mean ROSAT PSPC countrate of AG Dra during the all-sky survey was determined to (0.99±0.15) cts/sec. Similar countrates were detected in several PSPC pointings during the quiescent time interval 1991–1993. The X-ray light curve (mean countrate over each pointing) of AG Dra as deduced from the All-Sky-Survey data taken in 1990, and 11 *ROSAT* PSPC pointings as well as 7 HRI pointings taken between 1991 and 1996 is shown in Fig. 3. The countrates of the HRI pointings have been converted with a factor of 7.5 and are also included in Fig. 3. This 5 yrs X-ray light curve displays several features:

1. The X-ray intensity has been more or less constant between 1990 and the last observation (May 1993) before the optical outburst. Using the mean best fit blackbody model with kT = 15 eV and the galactic column density $N_H = 3.15 \times 10^{20}$ cm$^{-2}$ (see below), the unabsorbed intensity in the ROSAT band (0.1–2.4 keV) is $2.5 \times 10^{-9}$ erg cm$^{-2}$ s$^{-1}$.
2. During the times of the optical outbursts the observed X-ray flux drops substantially. The observed maximum amplitude of the intensity decrease is nearly a factor of 100. Due to the poor sampling we can not determine whether the amplitudes of the two observed X-ray intensity drops are similar. With the lowest intensity measurement being an upper limit, the true amplitude is certainly even larger.
3. Between the two X-ray minima the X-ray intensity nearly reached the pre-outburst level, i.e. the relaxation of whatever parameter caused these drops was nearly complete. We should, however, note that the relaxation to the pre-outburst level was faster at optical wavelength than at X-rays, and also was faster in the B band than in the U band. In December 1994 the optical V brightness ($\approx 9^m5$) was nearly back to the pre-outburst magnitude, while the X-ray intensity was still a factor of 2 lower than before the outburst.
4. The quiescent X-ray light curve shows two small, but significant intensity dips in May 1992 and April/May 1993. The latter and deeper dip coincides with the orbital minimum in the U lightcurve (Meinunger 1979, Skopal 1994).

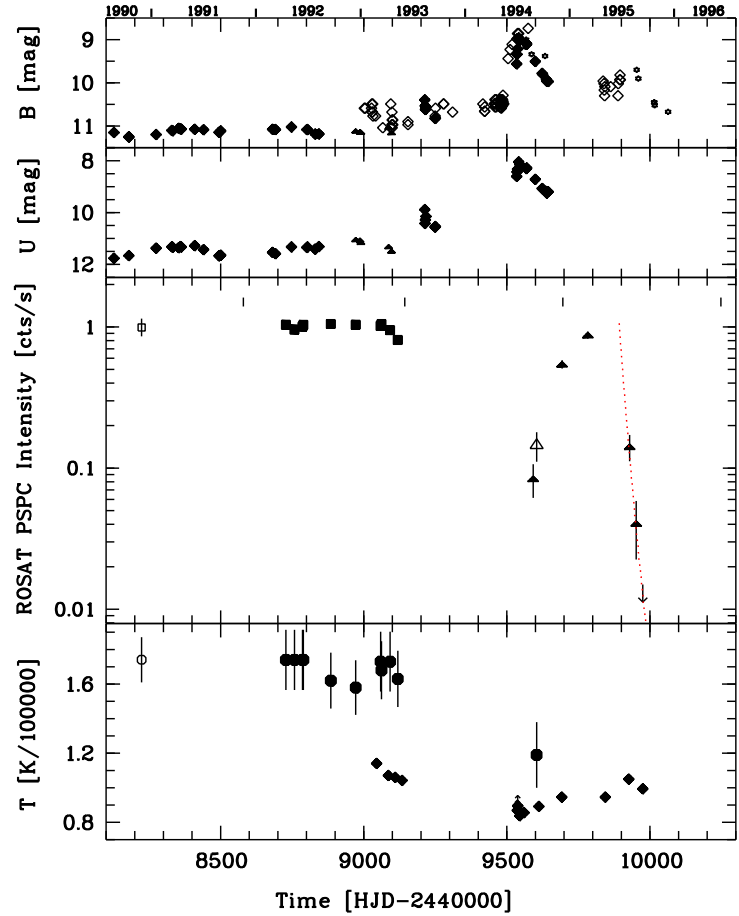

**Fig. 3.** The X-ray and optical light curve of AG Dra over the past 5 years. The two top panels show the U and B band variations as observed at Sonneberg Observatory (filled losanges = photoelectric photometry, open losanges = photographic sky patrol), Skalnate Pleso Observatory (Hric et al. 1994; filled triangles), and italian amateurs (Montagni et al. 1996; stars). The large middle panel shows the X-ray intensity as measured with the *ROSAT* satellite: filled squares denote PSPC observations, filled triangles are HRI observations with the countrate transformed to PSPC rates (see text), and the open triangle is the Boron filter observation corrected for the filter transmission. Statistical errors ($1\sigma$) are overplotted; those of the PSPC pointings are smaller than the symbol size. The vertical bars at the top indicate the minima of the U band lightcurve (Skopal 1994). The dotted line shows the fit of an expanding and cooling envelope to the X-ray decay light curve. The lower panel shows temperature estimates from blackbody fits to the *ROSAT* X-ray data (filled circles) and from the HeII($\lambda$1640) flux to continuum flux at $\lambda$1340 Å as determined from the IUE spectra and assuming E(B−V)=0.06 (=Zanstra temperatures, filled losanges).

### 3.3 The X-ray Spectrum of AG Dra in Quiescence

For spectral fitting of the all-sky-survey data the photons in the amplitude channels 11–240 (though there are almost no photons above channel 50) were binned with a constant signal/noise ratio of $9\sigma$. The fit of a blackbody model with all parameters free results in an effective temperature of $kT_{bb} = 11$ eV.

We investigated the possibility of X-ray spectral changes with time. First, we kept the absorbing column fixed at its galactic value and determined the temperature being the only fit parameter. We find no systematic trend of a temperature decrease (lower panel of Fig. 3). Second, we kept the temperature fixed (at 15 eV in the first run and at the best fit value of the two parameter fit in the second run) and checked for changes in $N_H$, again finding no correlation. Thus, no variations of the X-ray spectrum could be found along the orbit.

The independent estimate of the absorbing column towards AG Dra from the X-ray spectral fitting indicates that the detected AG Dra emission experiences the full galactic absorption. While fits with $N_H$ as free parameter systematically give values slightly higher than the galactic value (which might lead to speculations of intrinsic absorption), we assess the difference to be not significant due to the strong interrelation of the fit parameters given the energy resolution of the PSPC and the softness of the X-ray spectrum. We will therefore use the galactic $N_H$ value ($3.15 \times 10^{20}$ cm$^{-2}$ according to Dickey and Lockman 1990) in the following discussion.

With fixed $N_H$ the mean temperature during quiescence is about 14–15 eV, corresponding to 160000–175000 K. These best fit temperatures are plotted in a separate panel below the X-ray intensity (Fig. 3). The variations in temperature during the quiescent phase are consistent with a constant temperature of the hot component of AG Dra.

### 3.4 The X-ray Spectrum of AG Dra in Outburst

As noted already earlier (e.g. Friedjung 1988), the observed fading of the X-ray emission during the optical outbursts of AG Dra can be caused by either a temperature decrease of the hot component or an increased absorbing layer between the X-ray source and the observer. In order to evaluate the effect of these possibilities, we have performed model calculations using the response of the *ROSAT* HRI. In a first step, we assume a 15 eV blackbody model and determine the increase of the absorbing column density necessary to reduce the *ROSAT* HRI countrate by a factor of hundred. The result is a factor of three increase. In a second step we start from the two parameter best fit and determine the temperature decrease which is necessary to reduce the *ROSAT* HRI countrate at a constant absorbing column ($3.15 \times 10^{20}$ cm$^{-2}$). We find that the temperature of the hot component has to decrease from 15 to 10 eV, or correspondingly from 175000 K to 115000 K.

The only ROSAT PSPC observation (i.e. with spectral resolution) during the optical outburst is the one with Boron filter. The three parameter fit as well as the two parameter fit give a consistently lower temperature. But since the Boron filter cuts away the high-end of the Wien tail of the blackbody, and we have only 19 photons to apply our model to, we do not regard this single measurement as evidence for a temperature decrease during the optical outburst.

What seems to be excluded, however, is any enhanced absorbing column during the Boron filter observation. The best fit absorbing column of the three parameter fit is $4.4 \times 10^{20}$ cm$^{-2}$, consistent with the best fit absorbing column during quiescence. Since the low energy part of the spectrum in the PSPC is not affected by the Boron filter except a general reduction in efficiency by roughly a factor of 5, any increase of the absorbing column would still be easily detectable. For instance, an increase of the absorbing column by a factor of two (to $6.3 \times 10^{20}$ cm$^{-2}$) would absorb all photons below 0.2 keV and would drop the countrate by a factor of 50 contrary to what is observed.

## 4  Discussion

### 4.1  Quiescent UV Emission

UV observations show that the hot components of symbiotic stars are located in the same quarters of the Hertzsprung-Russell diagram as the central stars of planetary nebula (Mürset et al. 1991). Due to the large binary separation in symbiotic systems the present hot component (or evolved component) should have evolved nearly undisturbed through the red giant phase. However, presently the outermost layers of the white dwarf might be enriched in hydrogen rich material accreted from the cool companion.

The far-UV radiation of the WD is ionizing a circumstellar nebula, mostly formed (or filled in) by the cool star wind. The UV spectrum in quiescence is the usual in symbiotic systems: a continuum increasing toward the shortest wavelengths with strong narrow emission lines superimposed. The most prominent lines in the quiescence UV spectrum of AG Dra are, in order of decreasing intensity: HeII 1640 Å, CIV 1550 Å, NV 1240 Å, the blend of SiV and OIV] at 1400 Å, NIV] 1486 Å and OIII] 1663 Å. The continuum becomes flatter longward approximately 2600 Å due to the contribution of the recombination continuum originated in the nebula. Although both continuum and emission lines have been found to be variable during quiescence (e.g. Mikolajewska et al. 1995), there is no clear relation with the orbital period of the system. The ratio intensity of the recombination He II 1640Å line to the far-UV continuum at 1340 Å suggests a Zanstra temperature of around $1.0$-$1.1 \times 10^5$ K during quiescence (see Fig. 3).

### 4.2  Quiescent X-ray Emission

**The X-ray spectrum** Previous analyses of X-ray emission of symbiotic stars have been interpreted either with blackbody emission from the hot component (Kenyon and Webbink 1984, Jordan et al. 1994) or with bremsstrahlung emission from a hot, gaseous nebula (Kwok and Leahy 1984). Our ROSAT PSPC spectra of AG Dra show no hint for any hard X-ray emission. The soft spectrum is well fitted with a blackbody model. A thermal bremsstrahlung fit is not acceptable to this soft energy distribution. There is also no need for a second component as proposed by Anderson et al. (1981).

With an adopted distance of 1 kpc and using the blackbody fit parameters while $N_H$ was fixed at its galactic value, the unabsorbed bolometric luminosity of the hot star during quiescence (1990–1993) is $1.4 \times 10^{36}$ $(D/1\ {\rm kpc})^2$ erg/s (or equivalently 370 $(D/1\ {\rm kpc})^2$ $L_\odot$) with an uncertainty of a factor of a few due to the errors in the absorbing column and the temperature. The blackbody radius is derived to be $R_{bb} = 1.33 \times 10^9$ cm $(D/1\ {\rm kpc})$.

**Wind Mass Loss and Accretion** Accepting the high luminosity during quiescence and consequently assuming that the hot component in the AG Dra system is in (or near to) the hydrogen burning regime, the companion has to supply the matter at the high rate of consumption by the hot component. The spectral classification of the companion in the AG Dra binary system, in particular its luminosity class, is very uncertain. The commonly used K3III classification would imply a mass loss rate of $(0.2–7) \times 10^{-10}$ $M_\odot$/yr according to the formula of Reimers (1975) or that of De Jager et al. (1988) for population I objects. In contrast, a K0 supergiant could have a mass loss rate of about $5 \times 10^{-8}$ $M_\odot$/yr, more appropriate for the burning state of the compact object in AG Dra.

While a K3III giant of 1.1 $M_\odot$ would not fill its Roche lobe, a K0Ib supergiant would by far overfill its Roche lobe for any WD mass. Thus, in the case of a K3III companion the mass transfer to the WD would have to occur via Bondi-Hoyle accretion from the wind of the giant which by itself is already too low by orders of magnitude to sustain a burning compact object. Alternatively, a supergiant could either supply a high mass loss for wind accretion or could even feed an accretion disk around the WD by Roche lobe overflow.

We therefore think that in order to sustain the high mass comsumption of the compact object in AG Dra, the companion must be at least a bright giant to fill its Roche lobe. Adopting a bright giant solution, the distance to AG Dra would be of the order of 4 kpc, and thus the luminosity would be $2.2 \times 10^{37}$ erg/s. The implied compact object mass would be 0.4 $M_\odot$ (assuming the core mass-luminosity relation $L/L_\odot \approx 4.6 \times 10^4$ ($M_{core}/M_\odot - 0.26$) according to Iben and Tutukov 1989), and the mass ratio 13.

### 4.3  X-ray Emission During the Outbursts

Using our finding of an anticorrelation of the optical and X-ray intensity and the lack of considerable changes in the temperature of the hot component during the 1994/95 outburst of AG Dra, we propose the following rough scenario. (1) The white dwarf is already burning hydrogen stably before the outburst. (2) Increased mass transfer, possibly episodic, from the cool companion results in a slow expansion of the white dwarf. (3) The expansion is restricted either due to the finite excess mass accreted or by the wind driven mass loss from the expanding photosphere. This wind possibly also suppresses further accretion onto the white dwarf. The photosphere is expected to get cooler with increasing radius.
(4) The white dwarf is contracting back to its original state once the accretion
rate drops below the burning rate. Since the white dwarf is very sensitive it is not
expected to return into a steady state immediately (Paczynski and Rudak 1980).
Instead, it might oscillate around the equilibrium state giving rise to secondary
or even a sequence of smaller outbursts following the first one.

The scenario of an expanding white dwarf photosphere due to the increase in
mass of the hydrogen-rich envelope has already been proposed by Sugimoto et al.
(1979). The expansion velocity was shown to be rather low (Fujimoto 1982). If
we assume that the luminosity remains constant during this expansion, and for
the moment also assume that the change in the envelope mass is small during
the early phase (justified by the result; see below) then we can determine the
expansion velocity (or equivalently the excess accretion rate $\dot{M}-\dot{M}_{\rm RG}$) simply by
folding the corresponding temperature decrease by the response of the ROSAT
detector. Fitting the countrate decrease of a factor of 3.5 within 23 days (from
0.14 cts/s on HJD 9929 to 0.04 cts/s on HJD 9952) we find that

$$\left(\frac{\dot{M}-\dot{M}_{\rm RG}}{10^{-6} M_\odot\, yr^{-1}}\right) \approx 0.27$$

i.e. the necessary change in the accretion rate is as small as only 27%. This
translates into an expansion rate of $\approx$2 m/s. The corresponding modelled coun-
trate decrease is shown as the dotted line in Fig. 3, and the extrapolation to the
quiescent countrate level gives an expected onset of the expansion at JD $\approx$ 9891.

Since we did not observe the X-ray intensity decline during the first outburst,
and have no appropriate data yet for the X-ray intensity rise after the second
outburst, the following estimates are only crude. Assuming similar durations of
the first and the second outburst, we find that during each event the radius of the
white dwarf roughly doubles (from 0.02 $R_\odot$ to 0.043 $R_\odot$) while the temperature
decreases by about 35% (from 14.8 eV to 9.5 eV for the two parameter fit).

### 4.4   Relation to the Class of Supersoft X-ray Sources

With its soft X-ray properties AG Dra is very similar to the supersoft X-ray
sources (SSS) which are characterized by very soft X-ray radiation (kT$\approx$ 25–
40 eV) of high luminosity (Greiner et al. 1991, Heise et al. 1994). Among the
optically identified objects there are several different types of objects: close bi-
naries like the prototype CAL 83 (Long et al. 1981), novae, planetary nebulae,
and symbiotic systems. In addition to AG Dra, two other luminous symbiotic
systems are known (RR Tel, SMC 3) which both are symbiotic novae.

The similarity in the quiescent X-ray properties of AG Dra to those of the
symbiotic novae RR Tel and SMC 3 might support the speculation that AG
Dra is a symbiotic nova in the post-outburst stage for which the turn-on is not
documented. We have checked some early atlasses such as the Bonner Durch-
musterung, but always find AG Dra at the 10–11$^{\rm m}$ intensity level. Thus, if AG

Dra should be a symbiotic nova, its hypothetical turn on would have occurred before 1855. We note, however, that there are a number of observational differences between AG Dra and RR Tel which would be difficult to understand if AG Dra were a symbiotic nova.

If AG Dra is not a symbiotic nova, it would be the first wide binary supersoft source (as opposed to the classical close binaries) the existence of which has been predicted by Di Stefano et al. (1996). These systems are believed to have donor companions more massive than the accreting white dwarf which makes the mass transfer unstable on a thermal timescale. Recent calculations have shown that there is a maximum mass ratio of about 7, above which no system survives. The mass ratio of 13 for AG Dra which results from adopting a luminosity class II companion is far above this upper limit and thus suggests that more work has to be done to improve our understanding of the system parameters in AG Dra and/or that of the mass transfer in high mass ratio binary systems.

## 5 Summary

The major results and conclusions can be summarised as follows:

- The X-ray spectrum in quiescence is very soft, with a blackbody temperature of about 15 eV.
- The quiescent bolometric luminosity of $1.4\times10^{36}$ $(D/1 \text{ kpc})^2$ erg/s suggests stable hydrogen burning in quiescence.
- The monitoring at X-rays and UV wavelengths did not yield any hints for a predicted eclipse during the U light minima.
- In order to sustain the high luminosity the companion is required to fill its Roche lobe, and consequently is assumed to be of luminosity class II. With this assumption, the distance of AG Dra would be about 4 kpc, the X-ray luminosity would suggest a low-mass white dwarf, and the mass ratio would exceed 10.
- During the optical outburst in 1994 the UV continuum increased by a factor of 10, the UV line intensity by a factor of 2 and the X-ray intensity dropped by at least a factor of 100. There is no substantial time lag between the variations in the different energy bands.
- There is no hint for an increase of the absorbing column during one X-ray observation with spectral resolution performed during the decline of the optical outburst in 1994. Instead, a temperature decrease is consistent with the X-ray data and also supported by the *IUE* spectral results.
- Modelling the X-ray intensity drop by a slowly expanding white dwarf with concordant cooling, we find that the necessary excess accretion rate is only 27% of the quiescent one. Accordingly, the white dwarf expands to approximately its double size within the about three months rise of the optical outburst. The cooling is moderate: the temperature decreases by only ~35%.
- AG Dra could either be a symbiotic nova with a turn on before 1855, or the first example of the wide binary supersoft source class.

*Acknowledgement:* We are grateful to J. Trümper and W. Wamsteker for granting the numerous target of opportunity observations with *ROSAT* and *IUE* . JG is supported by the Deutsche Agentur für Raumfahrtangelegenheiten (DARA) GmbH under contract FKZ 50 OR 9201. RV is partially supported by the Italian Space Agency under contract ASI 94-RS-59. The *ROSAT* project is supported by the German Bundesministerium für Bildung, Wissenschaft und Forschung (BMBF/DARA) and the Max-Planck-Society.